\documentclass[preprint,aps,amsmath,amssymb]{revtex4-2}
\usepackage{graphicx}
\usepackage{bm}
\usepackage[usenames, dvipsnames]{color}
\usepackage[utf8]{inputenc}
\usepackage[T1]{fontenc}
\usepackage{mathptmx}
\usepackage{hyperref}
\usepackage{subcaption}
\usepackage{caption,setspace}
\usepackage{multirow}
\usepackage{float}
\usepackage{booktabs}
\usepackage{amsmath}
\usepackage[figuresright]{rotating}
\usepackage{soul}

\renewcommand{\figurename}{}
\renewcommand{\thefigure}{ Fig. \arabic{figure}}

\begin{document}

\title{Super-Arrhenius temperature-dependent viscosity due to liquid-liquid phase separation in the super-cooled Kob-Andersen model}
\author{Jayme Brickley}
\author{Xueyu Song}
\email{Corresponding author:xsong@iastate.edu}
\affiliation{Department of Chemistry and Ames Laboratory , Iowa State University, Ames, IA 50011, USA}

\begin{abstract}
In this study, a recently introduced order parameter called the weighted coordination number (WCN) was used to investigate the liquid-liquid (LL) phase separation, indicating temperature-dependent coarsening of the LL interface as a possible mechanism for the glass transition. A well-established glass-forming Kob-Andersen binary Lennard-Jones system was used in this study. The gas-liquid binodal line was reconstructed using WCNs, and the same approach was extended to study the liquid-liquid binodal line. Systems of various densities are instantaneously quenched from high to low temperatures where liquid-liquid separation is observed. The densities and composition of each liquid state were used to verify the level rule, along with the density and pressure profiles, demonstrating the local equilibrium of liquid-liquid phase separation.  The transition from the liquid-liquid phase separation in the supercooled region to the glass transition region was modeled by adopting a Markov Network Model to estimate the temperature-dependent viscosity using liquid-liquid interfacial information from the classification.
\end{abstract}
\maketitle

\section {INTRODUCTION}

Liquid-liquid phase transition (LLT) is a widely studied phenomenon, both experimentally and theoretically. The direct observation of LLT has occurred in many systems, such as phosphorus \cite{katayama2000first}. Simulated water systems have also provided convincing evidence for one or multiple LLT's~\cite{sciortino2011study,yagasaki,Liu,Singh}. However, the nature and universality of these transitions remain unclear. Density-driven heterogeneity readily allows for the study of gas-liquid or liquid-liquid equilibria using Gibbs Ensemble Monte Carlo or direct coexistance with pre-prepared protocols\mbox{~\cite{Gibbs, direct}}. The former method requires splitting the simulation system into two boxes, the latter requires carefully prepared systems near equilibria, hence disallowing direct observation of a separation. One can take advantage of structural heterogeneities to visualize liquid-solid coexistence via symmetry-based methods such as q4q6\mbox{~\cite{Viet-1}}, but a scheme that allows characterization and visualization of simple liquid-liquid equilibria is not available. 

For over a decade, studies have probed the free-energy landscape near the liquid-liquid critical point to establish evidence of such a first-order phase transition. For exmaple, this has been accomplished for many water models, including ST2, TIP4P/2005(ICE), and more recent machine-learning assisted models\mbox{~\cite{palmer2014,sciortino2024,poole2013,sciortino2025}}. While these free-energy methods yield undeniable evidences of two metastable liquid in coexistence, they lack the ability to directly observe and measure the LLT properties for large systems under various conditions. For exmaple, recent work by Wang $et. al.$ utilized their own order parameter to study high/low density liquid coexistence using iAMOEBA potential \mbox{~\cite{wang2026}}. While they were successful in observing HDL/LDL doamins, they were limited to small systems and used prepared configurations with HDL/LDL already in contact. The liquid-liquid separation and free energy barrier between domains disappeared for systems larger than 1,000 molecules. Additionally, while interfacial particles of these domains could be tracked, no thermodynamic analysis was presented for corresponding liquid states. It is clear that while such efforts have proved fruitful in certain situations,  yet comprehensive classification method for the study of liquid-liquid coexistences is not available before our work~\cite{Viet-1, Viet-2}. Recognizing this limitation, we wish to develop a method that allows direct classificaiotn and sampling of the structure and dynamics of both the liquid-liquid coexistence and the corresponding interface, the current work is a continuation of that effort.

In two recent studies, the weighted coordination number of particles was used as an order parameter to characterize the domains of a model supercooled liquid \cite{Viet-1, Viet-2}.
The salient feature of such an order parameter is its ability to differentiate the structural features of the macroscopic states in the structural space. Simultaneously, the particles belonging to different macroscopic states form intertwining domains in the configurational space by utilizing the particle's identity in both spaces. Using such an order parameter as a "classification indicator," we show that there is a clear liquid-liquid phase separation by measuring various thermodynamic and dynamic properties of the super-cooled Kob-Andersen model system. Our studies also show that these domain structures as a result of the liquid-liquid phase separation naturally lead to two types of relaxation dynamics: intra-domain relaxation largely due to diffusion inside a domain, and inter-domain relaxation, which is related to the coarsening kinetics of the first-order phase transition.

Of particular interest is the interplay between the dynamic slowdown of glass-forming liquids in the presence of  liquid-liquid separation. A key feature of glass-forming liquids is the presence of numerous dynamically heterogeneous regions at low temperatures. Das and Sastry have done extensive work uncovering the nature of dynamical crossover near the glass transition for the 80:20 Kob-Andersen Model\mbox{~\cite{DAS2022100098}}, corroborating the idea of many dynamical heterogeneous regions below the Mode-Coupling Temperature T$_{MCT}$=0.435. While many such domains may exist within a supercooled amorphous system, the interplay between the thermodynamically driven liquid-liquid phase separation and the dynamic heterogeneity of glasses remains unclear. It has been shown previously~\cite{Viet-2} that despite the heterogeneities present, the dynamical and thermodynamic properties of such a system are adequately described by two liquid states. If the presence of  LLT governs the dynamics of the glass, then the thermodynamics of LLT can be used to clarify the nature of glass transitions. 

In the current study, the phase diagram of such a liquid-liquid (L-L) phase transition is calculated, including the positions of the critical point from the L-L transition and the liquid-gas transition. More importantly, by computing the temperature-dependent viscosity in the supercooled region it was found that the coarsening kinetics of the L-L transition may provide a convincing picture of the glass transition in this model system. Because the roles of liquid-liquid phase separation in glass transitions have been extensively discussed from various perspectives in our recent paper~\cite{Viet-2}, this discussion is not repeated here.

The remainder of this paper is organized as follows. In the next section, our classification scheme, simulation details,  and the phase diagram of the model system are presented. In Section III, the density and pressure profiles of the L-L transition are characterized,  and from these profiles the surface tensions in different regions of the interface are calculated. In Section IV, using the surface tension of the system, a two-state Markov network model is used to calculate the temperature-dependent viscosity in the deep supercooled region, which agrees with the literature results. Some concluding remarks are presented in Section V.

\section{Classification Scheme}
\subsection{Simulation Details}
	In this study, simulations were performed using the binary Kob-Andersen Lennard-Jones potential strictly under the 80:20 mixture regime~\cite{Kob1, Kob2}. This model is well known for its ability to avoid crystallization well below the melting temperature, making it an excellent model system for studying viscous liquids and glass transitions. The specific ratio avoids crystallization of both type A and B particles~\cite{PhysRevLett.120.165501,10.1063/5.0237306} and is common for studing the glass transition. Liquid-liquid separations were observed in our previous study\cite{Viet-2} using this potential in addition to glass transition studies via equation of state constructions \cite{Ashwin-eos, Sastry-eos}. The Kob-Andersen model can be expressed as 
	
\begin{equation}
V(r)= 
\begin{cases}
 4\epsilon_{A,B} \left[ \left( \frac{\sigma_{A,B}}{r} \right)^{12} - \left( \frac{\sigma_{A,B}}{r} \right)^6 \right] & \text{for } (r\leq r_c)\\
 0 &  \text{for }  (r > r_c),
\end{cases}
\end{equation} 
where $\epsilon$ is the energy well depth, $\sigma$ is the atomic diameter and $r_{c}$ is the cutoff radius (set to $2.5\sigma_{AB}$). The system consisted of $80\%$ type A particles and $20\%$ type B particles, with the parameters $\sigma_{BB}/\sigma_{AA} = 0.88$, $\sigma_{AB}/\sigma_{AA} = 0.8$, $\epsilon_{BB}/\epsilon_{AA} = 0.5$, and $\epsilon_{AB}/\epsilon_{AA} = 1.5$. The model was asymmetric to allow particles of type B to act as agitators, preventing the system from crystallizing. Systems were run under the NVT or NPT ensemble with periodic boundary conditions while using the Nose-Hoover thermostat and barostat. The parameters for solid Ar were chosen for unit conversion, with $\sigma = 0.3405$nm , $m = 6.69$ x $10^{-26}$kg and $\frac{\epsilon}{k_{B}} = 119.8K$. The time step was set to $0.005\tau$ (12 fs), and the system sizes included $N=[16000,27000,32000]$. The liquid was prepared at a temperature of $T^{*}=1.2$, which is well above the mode-coupling temperature $T^{*}=0.435$ at $\rho^{*}=1.2$ estimated by the mode-coupling theory \cite{Testard}, and allowed to relax for one million time steps (12ns). Each system was then instantaneously quenched to the target temperature and equilibrated in the order of microseconds at lower temperatures, after which all relevant data were collected. 
	
\subsection{WCN as an order parameter for classification}

	It has previously been shown that weighted coordination numbers (WCNs) accurately describe the local structural and dynamic heterogeneities in supercooled liquids\cite{Viet-2}. In short, the method is based on commonly employed coordination numbers (CNs)  for normal liquids, with the addition of normalized Gaussian distributions at each notable solvation feature of the radial distribution function $g(r)$. Because the typical order parameters used to differentiate the states of matter fail for phases where all thermodynamic variables are similar such as condensed amorphous states, a new type of order parameters is needed to accurately describe a particle's local environment to capture the local structural differences of amorphous states. The highly averaged structural information given by $g(r)$ coupled with a more detailed description of each particle's neighboring environment allows for a balanced compromise to group particles with similar structural features. For the 80:20 binary KA mixture, the system is treated as an effective single component system due to prohibitive crystal growth and similar dynamical lengthscales for type A,B particles~\cite{kumawat2025}. As the liquid state is decreased further in temperature, dynamical and spatial heterogeneities begin to arise and individual CNs of particles increase dramatically with the liquid state complexity. WCNs alleviate the issue of individual assignment for each particle, rather than a statistical assignment with solvation features based on $g(r)$. This method provides an excellent balance between the strict assignment of CNs of individual particles, which leads to many structurally different states, and a single structural state based on $g(r)$.   
	
\begin{figure}[H]
\includegraphics[scale=0.4]{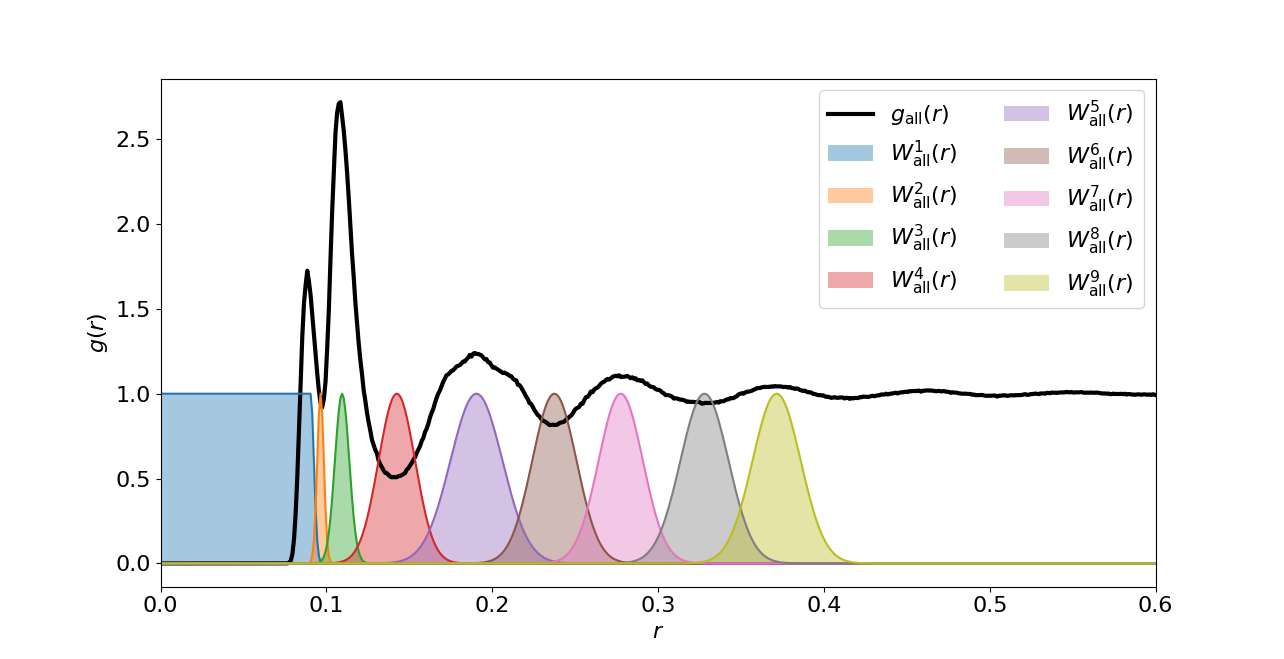}
\caption{The radial distribution function with weighted Gaussians placed at each notable feature for a KA binary LJ system at $T^{*}=0.43$, $\rho^{*} =1.12$. The Gaussian widths are chosen so that  the height at the intersection points is below 0.25.}
\label{fig:1}
\end{figure}
		
In this method, Gaussians are placed at the center of each feature of the total $g(r)$ and an intersection tolerance of $0$ to $0.25$ is allowed for the continuity of solvation shells. WCNs are treated as per-particle vectors with dimension equal to the number of features in $g(r)$, N=9 is shown in \ref{fig:1}. We have shown that the number of shells included in each WCN has no bearing on distinguishing the heterogeneities of the local structures once the major features of the $g(r)$ are captured.  Thus using more than nine features will not change our results. A particle's WCN is constructed by assigning a value to each particle's solvation feature by counting its neighboring particles weighted by that feature's Gaussian distribution~\cite{Viet-2}. These values are then summed and averaged for each feature such that $\overline{WCN_{i}} = \frac{1}{N_{b}}\sum_{j}^{N_{b}}WCN_{j}$, where $N_{b}$ is the number of neighboring particles $j$ of particle $i$ to obtain the final WCN of that particle after further smoothing.

\begin{figure}[H]
\resizebox{\columnwidth}{!}
{
\begin{subfigure}{0.45\textwidth}
\includegraphics[width=2.3in]{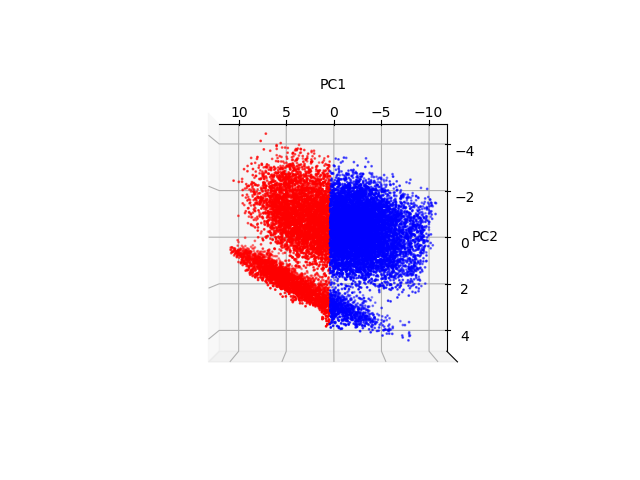}
\caption{2D projection in PC space}
\end{subfigure}
\hfill
\begin{subfigure}{0.45\textwidth}
\includegraphics[width=2.3in]{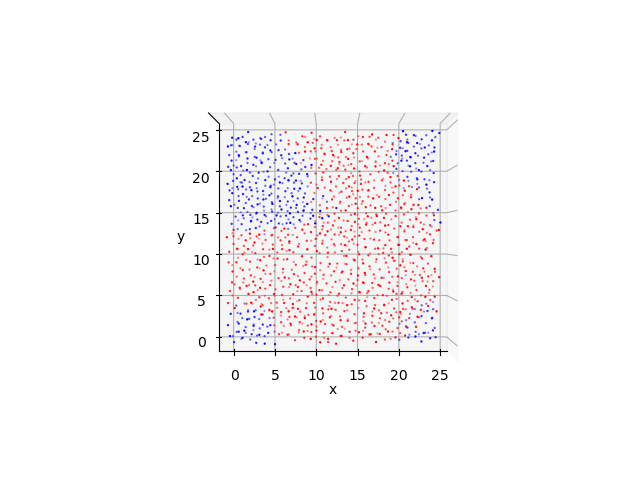}
\caption{2D projection in the configuration space}
\end{subfigure}
}
\parbox{6in}
{
\caption{ (a) 2D projection of PC space for KA binary LJ system at $T^{*}=0.3$, $\rho^{*} =1.12$. (b) Corresponding 2D projection in the configuration space using K-means clustering information  from PC space.}
}
\label{fig:2}
\end{figure}

\textit{N} features can then be represented by matrix \textbf{X} with dimensions $M\times N$ for a system of \textit{M} particles. Principal component analysis (PCA) is employed to reduce the number of features of \textbf{X}, which are reduced to a few highly correlated features for visual presentation. PCA is a useful non-parametric dimensionality reduction tool commonly used in data analysis with the following procedure:

a) Obtain mean-free data of WCNs $\tilde{X} = \bf{X} - \langle {\bf X} \rangle$, which is averaged over \textit{M} particles for each WCN.

b) Form the correlation matrix $C = \tilde{X}^T \tilde{X}$, which has the dimension $N$ x $N$. 

c) Solve the eigenvalue problem $Cu_{i} = \sigma_{i}^{2} u_{i}$ to obtain the principal components $u_i$. All PCs are retained and used to form the basis \textbf{U}, which is then dotted with the original dataset \textbf{X} to obtain the new feature space data $Y = U^TX$, which is called PC space.

 Although not required, PCA allows for a meaningful 3-dimensional view of the feature space. It is important to note that the success of the classification did not depend on the \textit{N} features being correlated. Systems of different morphologies often have distinct representations in the PC space which allows for the quick visual analysis of a system. This difference is most prominent in the arrangement of the PCs in the KA gas and liquid states, as shown in \ref{fig:3}. 
 
 We then use K-means clustering to extract clustering information from the correlated features in the feature space, setting $k=2$ with a priori knowledge that the system contains two distinct thermodynamic states (i.e. gas-liquid, liquid-liquid, etc.); larger $k$ values will converge to $k=2$ after the convergence test~\cite{Viet-2}. In short, higher initial k-values result in the shrinkage and gradual disappearance of all but two distinct liquid states upon iteration in PC space. K-means is an unsupervised,  geometric-based clustering algorithm that is excellent for distinguishing initial features in the PC space. The clustering information obtained by K-means can then be mapped from the PC space back to the real space given the identity of the particles. These clusters in real space serve as essential signatures for observing phase separation. 
 
 Depending on the thermal stability of the system interfacial particles must be treated with more care. Two different approaches are required to alleviate the hard assignment of interfacial particles from K-means in the real space and feature space. For the feature space, we employ the commonly used Gaussian Mixture Model which utilizes multivariate Gaussians and maximizes the log-likelihood for a softer assignment. In real space, we can tag interfacial particles and switch their cluster assignment, allowing clusters to grow or shrink in size. This co-learning strategy is iterated until the number of particles that change assignment is less than 1\% of the total system size, where convergence is documented in Ref.\cite{Viet-2}. Once converged, we gathered the relevant data for each cluster. Data can be collected for the density and pressure profiles using such an order parameter, and it is shown that there is no significant difference in these profiles after convergence. Hence this order parameter can be used to identify two thermodynamically different liquid states. The pressure and density profiles of the model system are described in Section~\ref{profiles}. 

Since we wish to observe liquid-liquid coexistence near the glass transition $T_G$, it is important to ensure observed states are thermodynamically metastable and not domianted by dynamical heterogeneities. Previous work by our group has demonstrated that domain growth of liquid states is conserved following the scaling law $L(t) \sim t^{1/3}$, and other quantities such as per-state diffusion coefficient ditributions and WCN distributions\mbox{~\cite{Viet-2}}. Additionally, we have also previously demonstrated the stability of liquid domain size after equilibration at low $T^*$ and is also shown here at $T^*=0.36$ (Appendix B).
 
It can be shown that the need for iterations to better define the liquid-liquid separation interface is dependent on the thermodynamics of the system. If thermal fluctuations are high, the need to iteratively define the interface is unnecessary, as the averages are thermally dominated. For lower temperatures near or below the glass transition temperature, the lack of thermal fluctuations calls for a more precise definition of the liquid-liquid interface. 
 
\subsection{Phase Diagram}

WCNs can distinguish any combination of states by characterizing small and large local structural differences, such as liquid-crystal~\cite{Viet-1}, liquid-liquid~\cite{Viet-2} and liquid-gas. 
To demonstrate this, the well established gas-liquid binodal line for the Kob-Andersen model is reconstructed and compared with those found in literature ~\cite{Testard}. The same strategy will be extended to the liquid-liquid binodal. Normally, liquid phases are indistinguishable from each other by conventional order parameters such as density. Therefore, the utility of WCN as an order parameter can be demonstrated by the construction of the gas-liquid and liquid-liquid phase diagrams within the entire thermodynamic variable space. 

\begin{figure}[H]
\begin{subfigure}{0.45\textwidth}
\includegraphics[width=2.3in]{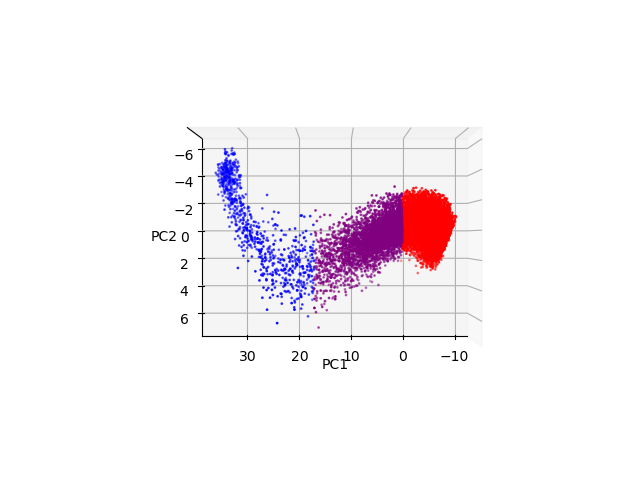}
\caption{WCNs in PC space for gas-liquid coexistence.}
\end{subfigure}
\hfill
\begin{subfigure}{0.45\textwidth}
\includegraphics[width=2.3in]{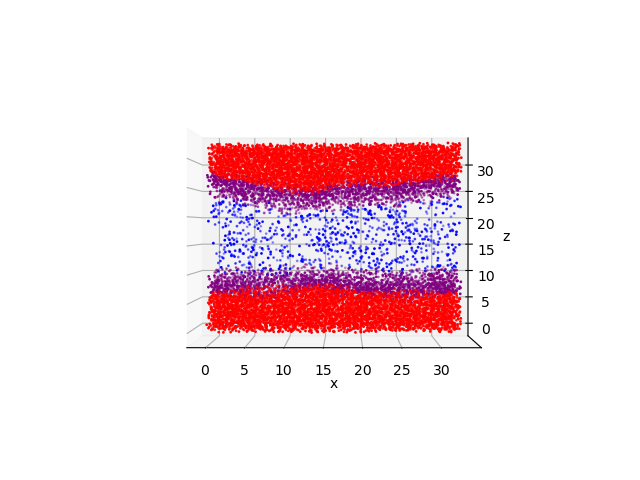}
\caption{Gas-liquid coexistence in configurational space}
\end{subfigure}
\caption{ (a) 2D projection of PC space for KA binary LJ system at $T^{*}=0.8$, $\rho^{*} =0.5$ with interfacial particles. (b) 2D projection of configuration space using K-means clustering information from PC space with k=3 to visualize interfacial particles. The natural k value in this case is k=2, as justified by thermodynamics of gas-liquid system. }
\label{fig:3}
\end{figure}

\begin{figure}[H]
\includegraphics[scale=0.4]{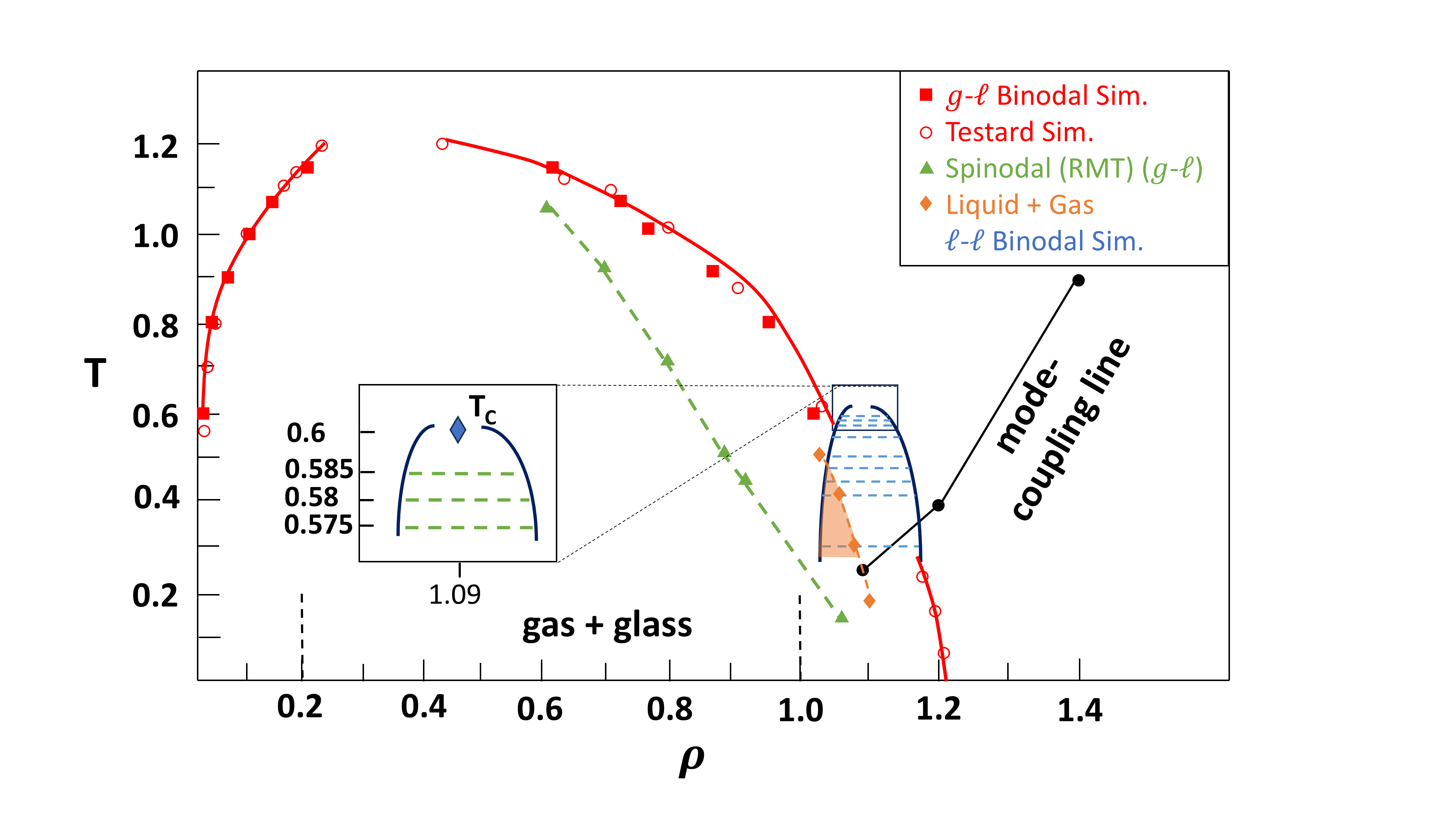}
\caption{Phase diagram of the KA binary LJ potential. All systems are equilibrated at $T^{*}=2.0$ before instantaneously quenching to the target temperature. The bulk density for gas-liquid systems was varied to obtain ideal separations, targeting spherical liquid clusters when possible. States were identified using WCN classification, which is then used to obtain average densities over 5ns. Densities at which the liquid-liquid separation was examined include $T^{*}=\{0.3,0.4,0.43,0.46,0.5,0.52,0.55,0.575,0.58,0.585\}$, and the lever rule was verified at key temperatures. 
Open circles are from \cite{Testard} and filled squares are from our simulations. The blue curve is the liquid-liquid binodal line from our simulations with detailed data listed in Table~\ref{table:binodal}.
The gas-liquid spinodal line (green), taken from ref \cite{Sastry-eos}, is estimated using Restricted Monte Carlo. Diamonds indicate end of gas nucleation and beginning of $l$-$l$ separation. Gas nucleation is thermodynamically favored over the liquid-liquid separation in the highlighted region, where only one liquid and a small gas bubble are observed. }
\label{fig:5}
\end{figure}

The calculated Kob-Andersen gas-liquid binodal line using the WCN order parameter was compared with a previous study~\cite{Testard}, which was constructed using density as an order parameter  in\ref{fig:5}. Each system was equilibrated at $T^{*}=2.0$ and quenched to target temperatures $T^{*}= \{ 1.15, 1.07, 1.0, 0.9, 0.8, 0.6\}$ with densities $\rho^{*} = \{0.4, 0.425, 0.445, 0.47, 0.5, 0.525\}$, respectively. After ten million time steps of equilibration, the averaged WC Ns were collected with $11$ frames spaced over a nanosecond. Gas and liquid particles are easily classified owing to the large difference in their local structures, driven by the density dependent $g(r)$. \ref{fig:3} demonstrates this for a planar interface at $T^{*}=0.8$, $\rho^{*} = 0.5$. The gas and liquid state information obtained from the classification was used to obtain the average density for every system, which was then used to construct the gas-liquid binodal line. Not only can WCNs differentiate states from one another, but they can also be used to show interfacial particles for a well-defined interface between a gas and liquid coexistence. In general, large structural or spatial heterogeneities are easily differentiated using WCNs. 

\begin{figure}[H]
\centering
\includegraphics[scale=0.4]{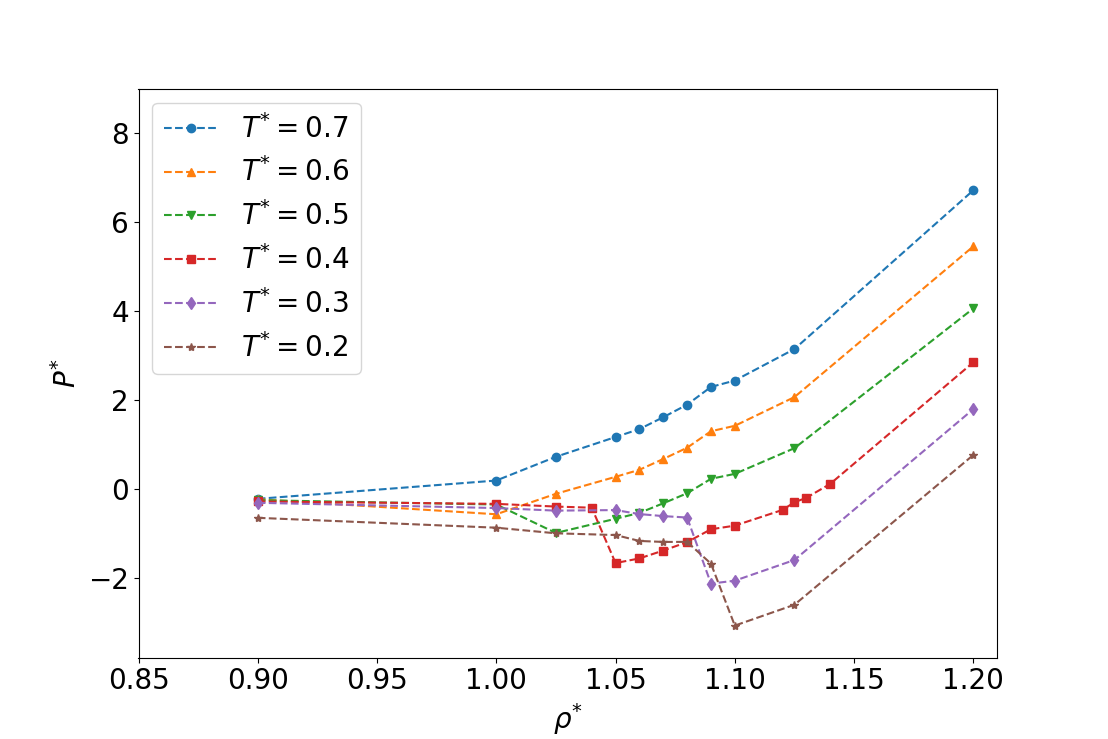}
\caption{Isotherms of the Kob-Andersen system. Monotonicity ends below $T^{*}=0.6$, signifying a first-order phase transition from liquid-liquid metastability to liquid-liquid unstablebility. Minima of non-monotonic isotherms, $\rho^{*}_{s}(T)$, represent densities in which gas nucleation ceases and liquid-liquid phase separation occurs.   }
\label{fig:4}
\end{figure}

Once the gas-liquid binodal line was obtained, we began the construction of the liquid-liquid binodal line. To determine where the gas phase falls out of equilibrium upon quenching from $T^{*}=1.2$, we computed isotherms to determine where the first-order phase transition from liquid-gas to liquid-liquid occurs. The restricted ensemble Monte-Carlo and empirical free energy methods used in Ref.\cite{Sastry-eos} to estimate the gas-liquid spinodal line yields a lower density limit for the gas phase instability, a difference of $\rho^*$= 0.05. Once the gas phase falls out of equilibrium with the liquid phase, metastability of the two liquid states is observed as shown in \ref{fig:4}. From our observations, there is competition in this region between gas formation and liquid-liquid separation. For isotherms $T^*\leq 0.5$, non-monotonicity is indicative of a liquid-gas to liquid-liquid transition. Bulk densities $\rho^* > 1.125$ result in no liquid-liquid formation within the given simulation limit. Similarly, points prior to non-monotonicity exhibit gas nucleation during cooling or during the 1.2$\mu$s of equilibration. Whether a liquid-liquid separation occurs prior to gas nucleation has not been studied, but this could be the case. Further study in this region might possibly yield the transition from liquid-liquid to liquid and subsequently the entire van der Waals style loop, which is beyond the scope of this study. The gas-liquid-liquid triple point was not observed in this study. 

These densities from the isotherms then serve as a lower limit to start scanning for the liquid-liquid phase transition while avoiding gas nucleation. By sampling various bulk densities and quenching to the target temperatures, the liquid-liquid binodal line can be obtained by measuring the densities of the classified states using WCNs. The densities of the liquid states were obtained by finding large clusters of each state within the simulation box following the classification, selecting a volume deep within the clusters, and monitoring the number of particles that remained within the trajectory. The equilibration time for the sampled state points ranged from 50-300 million timesteps (between 600ns and 3.6 microseconds) depending on the target temperature, and the densities were averaged over 5ns. The percentage of each liquid state that constitutes the simulation box, or $\chi_{i}$, is used to check the lever rule at each bulk density, which is a stringent test for thermodynamic equilibrium. According to the lever rule, the density of the two liquid states should not change with the bulk density at a given temperature and should only change with their partition within the system. Indeed, the ratio of the liquid states is in agreement with the bulk density (Table~\ref{table:level_rule}). This procedure is performed for $T^{*}=0.3$, $0.4$, and $0.5$. Both quenched and steps-cooled systems reach similar configurational arrangements and percentage compositions within the simulation box (Appendix B). 

\begin{table}[H]
\centering
\resizebox{8cm}{!}
{
\begin{tabular}{|l|l|l|l|l|}
\hline
$\rho^*_{bulk}$ & $\rho^*_1$ & $\rho^*_2$ & \% State 1 & \% State 2 \\ \hline
1.08       & 1.03 & 1.17 & 60              & 40              \\ \hline
1.09       & 1.03 & 1.17 & 52              & 48              \\ \hline
1.10       & 1.03 & 1.18 & 47              & 53              \\ \hline
1.12       & 1.03 & 1.18 & 40              & 60                \\ \hline
\end{tabular}
}
\caption{Table of quenched NVT systems at $T^{*}=0.4$, demonstrating the lever rule for each sampled bulk density. }
\label{table:level_rule}
\end{table}

By sampling various bulk densities and quenching to the target temperatures, the liquid-liquid binodal line can be obtained by measuring the densities of the classified states. The estimated liquid-liquid binodal line is shown in \ref{fig:5} and tabulated in Table~\ref{table:binodal}. The critical point can be estimated using the law of rectilinear diameters $(\rho_g + \rho_l)/2 = aT + b$ \cite{watanabe2012}. The critical density, $\rho_c^*=1.1$ is estimated based on the density data near the critical point. Three data points were used near the critical point, the parameters from the fitting lead to $a=0.92$ , $b=0.55$, hence the critical temperature $T_c^*=0.6$. Obtaining more accurate critical point data is beyond the scope of this paper, as testing finite size effects near $T^*_C$ is incredibly time consuming within the current aveWCN procedure.

\begin{table}[H]
\centering
\resizebox{12cm}{!}
{
\begin{tabular}{|l|l|l|l|l|l|l|l|l|l|l|}
\hline
$T^*$ & 0.3 & 0.4 & 0.43 & 0.46 & 0.5 & 0.52 & 0.55 & 0.575 & 0.58 & 0.585 \\ \hline
$\rho_1^*$ & 1.03 & 1.034 & 1.035 & 1.041 & 1.043 & 1.051 & 1.052 & 1.059 & 1.077 & 1.082 \\ \hline
$\rho_2^*$ & 1.194 & 1.189 & 1.159 & 1.147 & 1.140 & 1.138 & 1.134 & 1.122 & 1.118 & 1.112 \\ \hline
\end{tabular}
}
\caption{Binodal line of the liquid-liquid coexistence of the Kob-Andersen model. Systems are quenched to target temperatures from $T^{*}=2.0$, all at bulk density $\rho^* = 1.09$. Each system is equilibrated for a minimum of 600ns or maximum of 3.6$\mu$s depending on $T^*$, with data collection occurring over 12ns. Vertical slices are taken over various regions in each simulation box and averaged, using small areas where state information is stagnant over the course of the trajectory. Uncertainty values in density never exceed 0.01 reduced density unit.}
\label{table:binodal}
\end{table}

\section{Density and Pressure Profiles}
\label{profiles}
To demonstrate the mechanical equilibrium along the rough interface of liquid-liquid separation, relatively large regions of positive and negative curvatures must be thoroughly examined via pressure and density profiles to extract the underlying thermodynamic properties. Systems of size N=32000 were equilibrated in the NPT ensemble at zero pressure, $T^{*}=1.2$. The system was then quenched down to $T^{*}=0.3$ and equilibrated for $1.5$x$10^{8}\tau$. Per-atom pressure tensors were obtained via the LAMMPS~\cite{LAMMPS} molecular dynamics package using its per-atom stress tensor. An appropriate spherical interfacial region was identified for the high pressure state, and the six Cartesian stress tensor components were transformed into spherical coordinates with respect to the center of the cluster. Owing to the low magnitude of the off-diagonal terms, the pressure can be expressed as 
\begin{equation}
P(r) = P_{N}(r)e_{r}e_{r} + P_{T}(r)(e_{\theta}e_{\theta} + e_{\phi}e_{\phi}),
\end{equation}
where $\it{e_r}$, $\it{e_{\theta}}$, $\it{e_{\phi}}$ are the spherical unit vectors. $P_{N}$ and $P_{T}$ are the normal and tangential pressure components, respectively.  $P_{N}$ and $P_{T}$ are obtained via the numerical average of the pressure within the thin spherical shells extending from each cluster center. This procedure is performed for both types of curvature along the interface within the liquid-liquid separation domains. Density data were collected over the same bins. Two different types of curvatures exist across the liquid-liquid interface: convex(positive) and concave(negative). Both regions exhibited similar, but different,  pressure and density profiles. Working from the low pressure liquid state, density and pressure profiles can be obtained for both types of interfaces. This peculiar situation arises from the concave case, where negative curvature acts as destructive interference towards the total surface tension, wherein mechanical equilibrium is not maintained perfectly, an indication that the system is in a local equilibrium which leads to coarsening kinetics over a longer time scale~\cite{Viet-2}. 

Mechanical equilibrium is expressed as 
\begin{equation}
P_T (r) = P_N (r) + \frac{r}{2} \frac {dP_N (r)} {dr},
\end{equation}
which satisfies the condition $\bm{\nabla } \cdot \bm{P} = 0$~\cite{widom}. One of the challenges in studying these different curvature sections across an interface comes from the amount of available surface data, as the desired regions are rather small regardless the system size. Using the thin sheet approach, a given curved region of an interface may only have 40-50 particles per sampled shell, making statistically accurate averaging in these regions difficult at low $T^*$. 

\begin{figure}[H]
\begin{subfigure}{0.45\textwidth}
\includegraphics[width=2.3in]{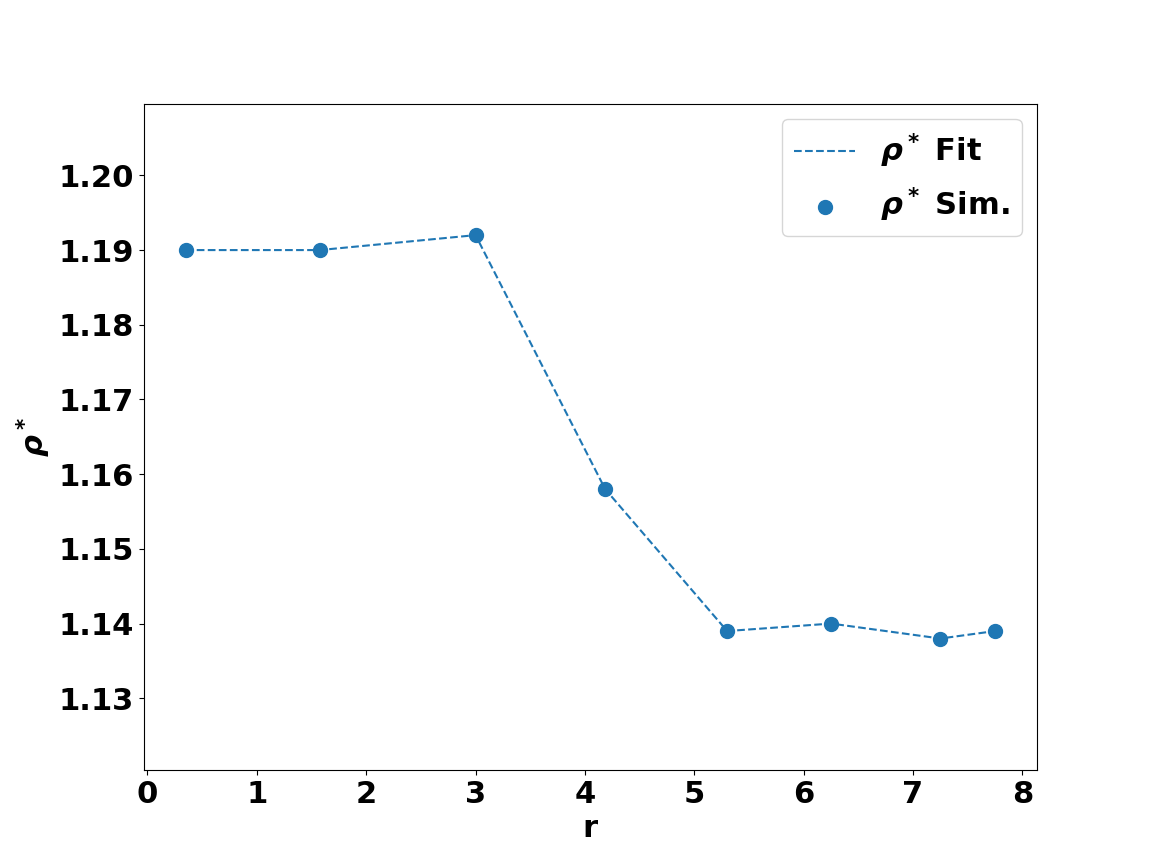}
\caption{Density Profile}
\end{subfigure}
\hfill
\begin{subfigure}{0.45\textwidth}
\includegraphics[width=2.3in]{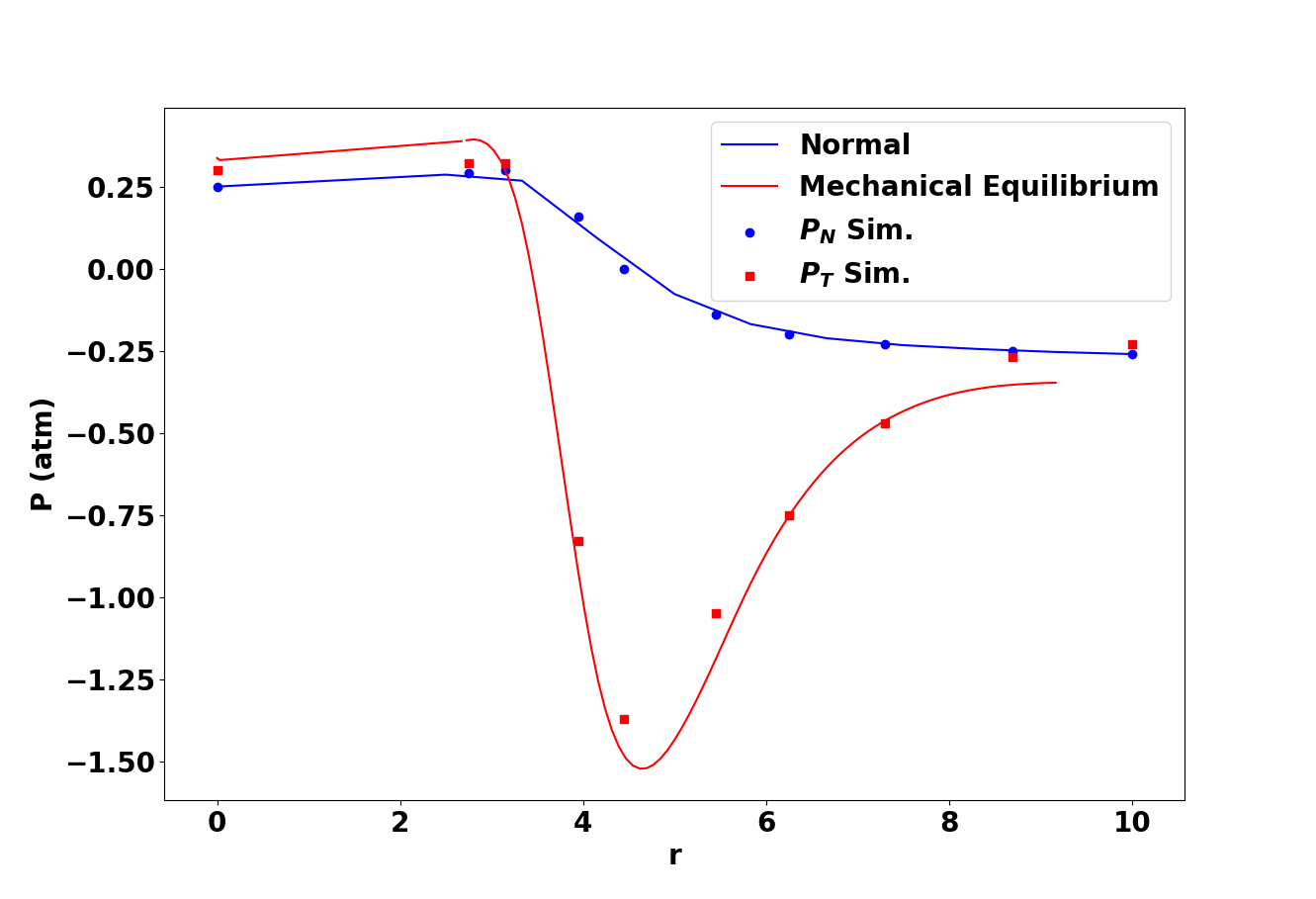}
\caption{Pressure Profile}
\end{subfigure}
\caption{ (a) Density profile and (b) pressure profile of $T^{*}=0.3$, $\rho^{*} =1.17$.  NPT system at zero pressure for convex region of the liquid-liquid interface. The surface of tension lies at $R_s=3.94$ (Table \ref{table:3}). Error for each data point $< 0.01$ \it{atm}.}
\label{fig:6}
\end{figure}

\begin{figure}[H]
\begin{subfigure}{0.45\textwidth}
\includegraphics[width=2.3in]{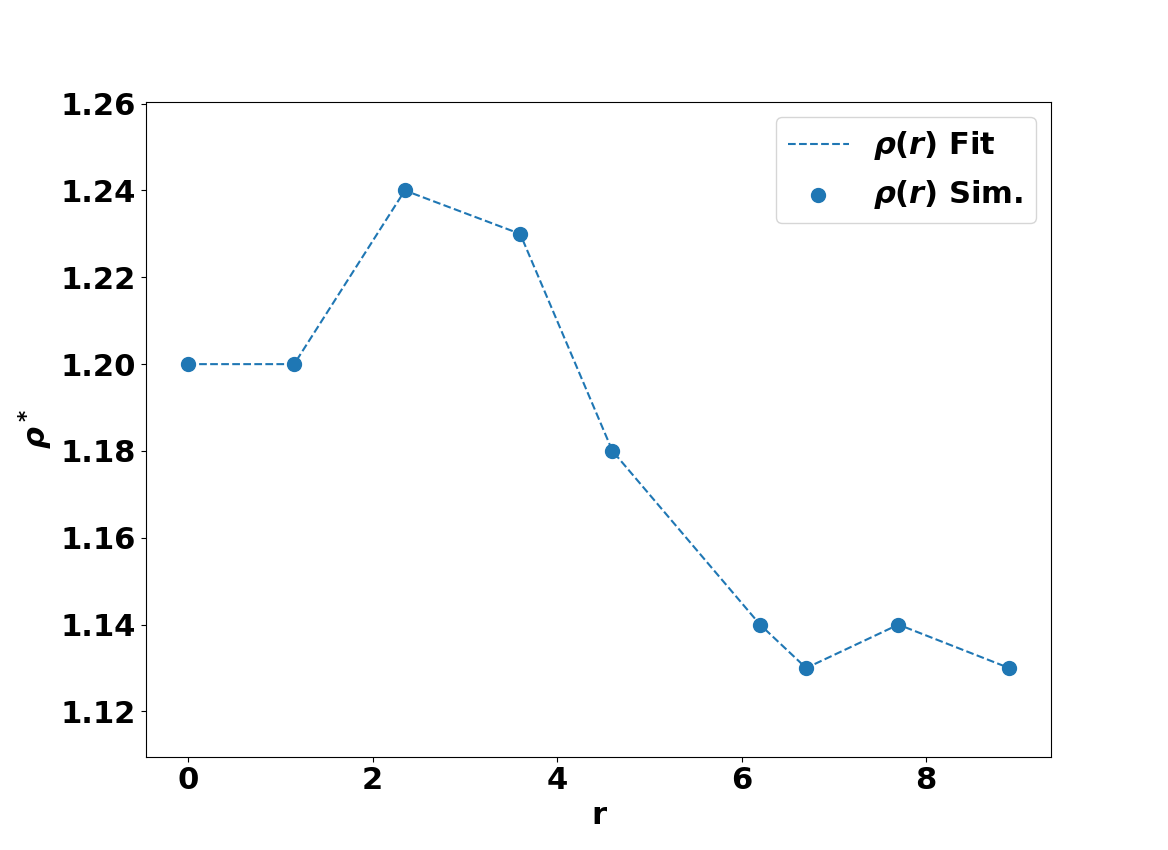}
\caption{Density Profile}
\end{subfigure}
\hfill
\begin{subfigure}{0.45\textwidth}
\includegraphics[width=2.3in]{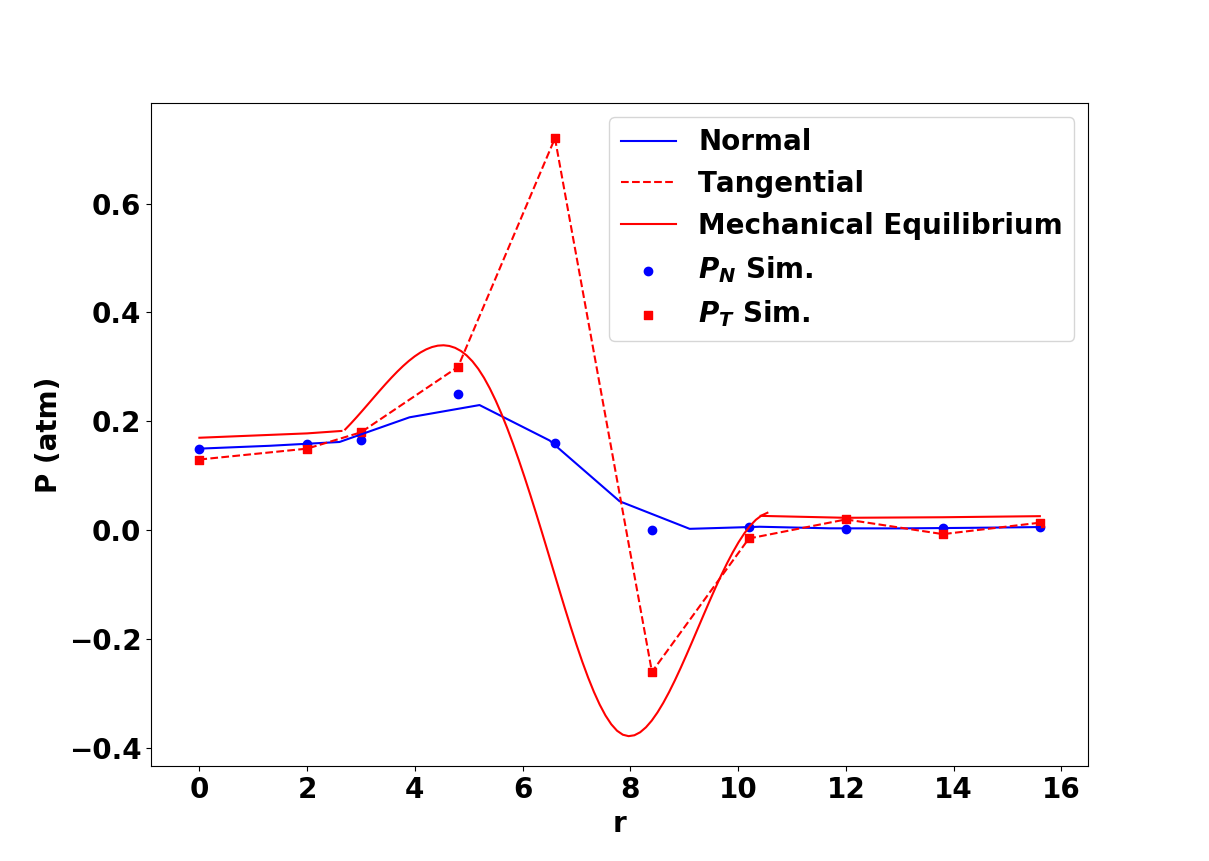}
\caption{Pressure Profile}
\end{subfigure}
\caption{ (a) Density profile and (b) pressure profile of $T^{*}=0.3$, $\rho^{*} =1.17$. NPT system at zero pressure for concave region of the liquid-liquid interface. The surface of tension lies at $R_s=3.84$ (Table \ref{table:3}). Error for each data point $< 0.009$ \it{atm}.}
\label{fig:7}
\end{figure}

To ensure proper sampling, block average within a given time interval can be obtained for each bin of particles~\cite{Frenkel-Smit} (See Appendix C). Block averaging is a common statistical analysis technique that  ensures that the data after a certain time \textit{t} are uncorrelated. The obtained pressure and density profiles paint a unique picture of the phase separation that maintains the mechanical equilibrium. The results obtained for the convex case (~\ref{fig:6}) mimic those found in most other convex fluid systems~\cite{Song};  however,  the concave pressure profile is uniquely different and minutely breaks the mechanical equilibrium. The data for the concave case were studied at two temperatures, $T^{*}=0.2,0.3$, which both show the same qualitative behaviors. Both the density and pressure profiles in the concave case (\ref{fig:7}) show an increase in $\rho$ and $P_{N}$ shortly before the surface of tension. This suggests local inhomogeneity along these regions of the liquid-liquid interface, and the breaking of the mechanical equilibrium provides a driving force for interface evolution at longer time scales. 

We can integrate the pressure and density profiles of the two curvature types, 
\begin{equation}
R_s =\frac{  \int_{r_a}^{r_b} r^2(P_N-P_T) \, dr }{  \int_{r_a}^{r_b} r(P_N-P_T) \, dr }
\end{equation}
and 
\begin{equation}
S_e =  \int_{r_a}^{r_b}\rho(r) ln\frac{\rho(r)}{\rho_0}dr,
\end{equation}
where $\rho_0$ is the overall density of the system and $r_{a,b}$ are the radial distances beginning in state $a$ and ending in state $b$. In our case, data collection was taken from the center of the cluster; therefore,  the integral $S_e$ was taken from $r_a=0$. The excess entropy, $S_e$ across the interface is negative in the convex case and positive in the concave case, which provides a measure of the surface tension at different regions of the interface and serves as a driving force for interface coarsening.  

The general method for computing the surface tension of a fluid system is to integrate the pressure profile,
\begin{equation}
\sigma_{s} = \int_{0}^{\infty} \left( \frac{r}{R_s} \right) ^{2}(P_N-P_T) \, dr,
\end{equation}
where $R_s$ is taken as the surface of tension and $\sigma_s$ from the integration of the pressure profiles can also be checked against the values obtained from the local curvature estimation, as detailed in the next section. Some of the data for each curvature case are listed in Table~\ref{table:3}. 

The surface tension as an excess free energy is related to the excess entropy via $\sigma_s = - S_eT$, as the excess energy can be neglected because of the small density change in the interfacial region~\cite{harsha}. Because the surfaces sampled to obtain the pressure profiles are larger than those of the entropy calculation, the excess surface free energy is scaled down proportionally to the $S_e$ interface. The excess entropy $S_e$ in the convex case is $S_e(convex) = -0.076$, while $S_e(concave) = 0.20$ at $T^{*}=0.3$, which compare well with $\sigma_s(convex) = 0.054$ and $\sigma_s(concave) = -0.065$ from Table~\ref{table:3}. Due to the physical ambiguity of negative surface tension, it is important to clarify that these regions are driven by sign-dependent local curvature. The density and pressure profiles of these concave regions of curvature seemingly break mechanical equilibrium. The exact physical nature of these regions is unclear and remains an area of research interest. 

\begin{table}[H]
\begin{tabular}{|l|l|l|l|l|}
\hline
Curvature & $R_s$(integration) & $R_c$(local) & $\sigma_s$(integration) & $\sigma_s$(local) \\ \hline
Convex       & 3.94 & 4.54 & 0.037             & 0.054              \\ \hline
Concave      & 3.84 & -5.68 & -0.085              & -0.065          \\ \hline
\end{tabular}
\caption{Values for the surface of tension, radius of surface curvature, and surface tension for convex and concave case for $T^{*}=0.3$ $\rho^{*}=1.17$, comparing values obtained from integration of pressure profiles to data obtained from local curvature fitting. (Note: Surface of tension $R_s$ and radius of curvature $R_c$ are not correlated physically.) }
\label{table:3}
\end{table}

After the initial quenching of the liquid-liquid phase separation in the supercooled region, the coarsening kinetics are driven by the reduction in the surface tension. Given the complex morphology of the interfaces, our curvature dependent surface tension calculations may present a concrete method for calculating viscosity as a function of temperature.

\section{Temperature dependent viscosity}

A hallmark of glass transition is the exponential growth of viscosity as the temperature is lowered in the supercooled region~\cite{wolynes_book}. Owing to the difference of many orders of magnitude between the viscosity near the melting temperature and the glass transition temperature, the traditional method to compute the viscosity, such as the  Green-Kubo method, cannot  be used across the entire temperature range because of the slow dynamics as the viscosity becomes many orders of magnitude larger than that of a normal liquid. 

To overcome this limitation, Yip et al.
developed a Markov network model to calculate the viscosity of slow dynamic systems~\cite{Li}. The model utilizes a variation of the Green-Kubo theory,  which describes stress relaxation using a Markov network of nodes. Each node contains a collection of particles called basin~\textit{i} with its own inherent stress 

\begin{equation}
\sigma_{i} = \frac{1}{\Omega} \langle -Nk_B T \bm{I} + \sum_{n=1}^{N} \ \bm{x}_n \bigotimes \partial_{\bm{x}_{n}} V \rangle_{i},
\end{equation}
where \textbf{I} is the 3x3 identity matrix. The two terms of the thermal average $\langle \rangle_{i}$ are the kinetic and virial components of the stress for the particles within each basin. The likelihood of a particle leaving basin \textit{i} for basin \textit{j} can be defined by the transition rate:
\begin{equation}
a_{ij} = \nu_0 e^{\frac{-q_{ij}}{k_B T}},
\end{equation}
where $\nu_0$ is the trial frequency and $q_{ij}$ is the activation barrier. Assuming that the averaged stress for each basin is a Markovian process, the shear stress time correlation function $\langle \sigma(t)\sigma(t + \tau) \rangle$ used for viscosity calculations within the linear response theory can be substituted  by a shear stress average over each basin, such that
\begin{equation}
\sigma(T) = \sum_{i} \ \sigma_i p_i(t),
\end{equation}
where $p_i(t)$ is a state-residence function for each basin, equal to one if the system is in basin \textit{i} at time \textit{t}, and zero otherwise. The Green-Kubo viscosity expression can be transformed as 

\begin{equation}
\eta (T) = \frac{\Omega}{k_BT} \int_{0}^{\infty}d\tau\langle\sigma(t)\sigma(t+\tau)\rangle=\frac{\Omega}{k_BT} \sum_{i} \ P_i \sigma_i \frac{1}{a_i} ({\bf{A}}(\omega=0^{+})^{-1} \sigma )_i,
\label{markov}
\end{equation}
where $\Omega$  is the volume, $P_i$ is the probability of belonging to basin $i$, $\sigma_i$ is the shear stress of basin $i$, $a_i$ is the transition rate to basin $i$. $({\bf{A}}(\omega=0^{+})^{-1} \sigma)_i=\sum_j({\bf{A}}(\omega=0^{+})^{-1})_{ij}\sigma_j$ is the propagator matrix between basin $i$ and all others. The propagator matrix overcomes the need for a shear stress time correlation function calculation of the entire system, as the slow dynamics of the system make averaging over each basin within the system a good approximation. 

In their application of the Markov network model to the Kob-Andersen model, they used the inherent structures of the potential energy surface as a starting point for various temperature sampling of different basins of the super-cooled system. A  crossover from an essentially Arrhenius scaling behavior at high temperatures to a low-temperature
behavior that is clearly super-Arrhenius is observed for the Kob-Andersen model~\cite{Li}.  In the glass transition temperature region,  thousands of basins are needed for sampling,  which makes physical interpretation difficult. 

 In our case, the liquid-liquid phase separation picture naturally leads to two different types of basins at near equilibrium conditions. The coarsening kinetics, which govern the particle transitions between different domains, are activated,  and the activation energy can be measured by the surface tension between the domains. If there are only two basins(domains) at quasi-equilibrium, then $P_1=P_2=1/2$, $\sigma_1=-\sigma_2=\sigma_0$, and $a_{12}=a_{21}=\nu_0\exp(-Q/k_BT)$, then Eq.(\ref{markov}) reduced to 
\begin{equation}
\eta (T) = \frac{\Omega}{k_BT} \sigma_{0}^{2} \nu_{0}^{-1} e^{\frac{Q}{k_BT}},
\label{eq:two_state_1}
\end{equation}
where $Q$ is the activation energy for particle transition between the two basins and, $\nu_0$ is the frequency at which the particles switch basins.

From our simulations the shear stresses between the two liquid states are equal and opposite. Symmetry of pressure and shear stress was confirmed for all temperatures used to construct viscosity. Eq.~\ref{eq:two_state_1} then becomes
\begin{equation}
\eta (T) = \frac{\Omega}{k_BT} \sigma_{0}^{2} \nu_{0}^{-1} e^{\frac{\phi}{k_BT}},
\label{eq:two_state_2}
\end{equation}
where $Q=\phi$ is the summation of the local surface tension of the liquid-liquid interface multiplied by the corresponding surface area, that is,  the total interfacial free energy.  The surface tension calculated in the previous section for positive and negative curvatures can act as constructive and destructive interference, which provides a physical mechanism for slow coarsening kinetics. In the previous section, the surface tension was calculated using the pressure profile at the interface; however, this is a time-consuming process because of the long average of the pressure profile calculations. As the system is nearly at mechanical equilibrium, the Laplace equation can be used for the calculation of the surface tension and it is demonstrated in Table~\ref{table:3} that the Laplace equation calculation yields essentially the same result as the pressure profile calculation.

Therefore, the per-particle local curvature is calculated for all interfacial particles in the system and the activation energy of the entire interface is
\begin{equation}
\phi= \frac{1}{2} \Delta P \sum_i \frac{a_i}{\kappa_i} = \frac{1}{2} \Delta P \overline{a} \sum_i \frac{1}{\kappa_i} ,
\label{eq:local}
\end{equation}
where $\Delta P$ is the difference in normal pressures between the two thermodynamic states, $a=\pi r^2$ is the surface area of the interfacial particles, and $\kappa_i$ is the local curvature of interfacial particle $i$. This local surface tension approach requires an accurate representation of the local curvature, which is obtained by generating a surface using opposite-state neighbor lists of interfacial particles. Due to the nature of generating these particles, we found it more appropriate to use the average area per-particle to calculate the activation energy of the interfeace (Eq. 13). The log-likelihood from the PC space can also be used to find interfacial particles, as detailed in Appendix~\ref{sec:local_curvature}. 

To ensure that the calculation of interfacial free energy, particularly $\sum_i \frac{a_i}{\kappa_i}$, is independent of the system size, viscosity was calculated using Eq.\ref{eq:two_state_2} for sizes $N=[8000, 15625, 19683]$. The summation of the interfacial surface area  times the inverse curvature,  $\sum_i \frac{a_i}{\kappa_i}$,  is independent of the system size $N$ as shown in Table~\ref{table:size_curvature} and \ref{fig:8} at $T^*=0.36$. 

\begin{table}
	\centering
	\begin{subtable}[t]{0.45\textwidth}
		\centering
		\begin{tabular}{|l|l|l|l|}
		\hline
		$N$ & $N_i$ & $\overline{a}\sum_i \frac{1}{\kappa_i}$ & $\overline{\kappa_i} $\\ \hline
		8000 & 1051 & 270 & 0.067 \\ \hline
		15625 & 1625 & 262 & 0.05 \\ \hline
		19683 & 3470 & 271 & 0.031 \\ \hline
		\end{tabular}
		\caption{$T^*=0.36$}
		\label{table:5a}
	\end{subtable}
	\begin{subtable}[t]{0.45\textwidth}
		\centering
		\begin{tabular}{|l|l|l|l|}
		\hline
		$N$ & $N_i$ & $\overline{a}\sum_i \frac{1}{\kappa_i}$ & $\overline{\kappa_i} $\\ \hline
		8000 & 941 & 232 & 0.084 \\ \hline
		15625 & 2208 & 241 & 0.070 \\ \hline
		19683 & 3343 & 249 & 0.033 \\ \hline
		\end{tabular}
		\caption{$T^*=0.38$}
		\label{table:5b}
	\end{subtable}
\caption{Relevant surface tension values per system size. $\sum_i \frac{a_i}{\kappa_i}$ is independent of $N$, while $N_i$, the number of interfacial particles, increases the mean curvature $\overline{\kappa_i}$ decreases.  As the system size increases, regions of negative and positive curvature both increase (\ref{fig:8}), the final $\sum_i \frac{a_i}{\kappa_i}$ remains as a constant within statistical error.}
\label{table:size_curvature}
\end{table}

Finite box size effects at lower $N$ values cause single cluster nucleation with a high curvature. As system size increases, the mean curvature $\overline{\kappa_i}$ decreases as the regions of negative curvature increase and both liquid states become bi-continuous, as shown in \ref{fig:8}. Therefore, the interfacial free energy remained constant as a function of the system size.  
\begin{figure}[H]
\begin{subfigure}[t]{0.3\textwidth}
\includegraphics[width=\textwidth]{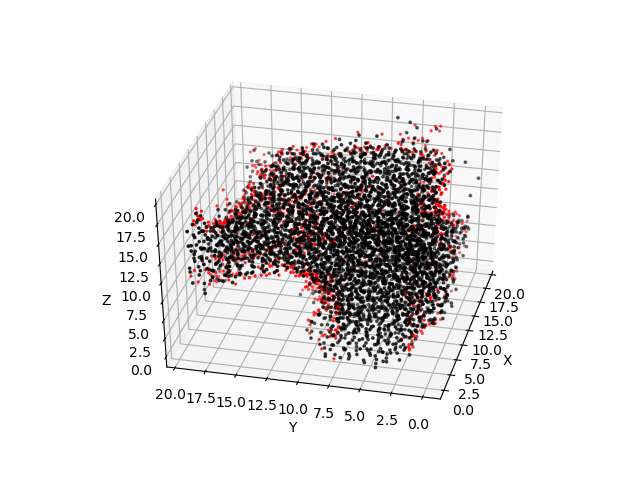}
\caption{N=8000}
\end{subfigure}
\begin{subfigure}[t]{0.3\textwidth}
\includegraphics[width=\textwidth]{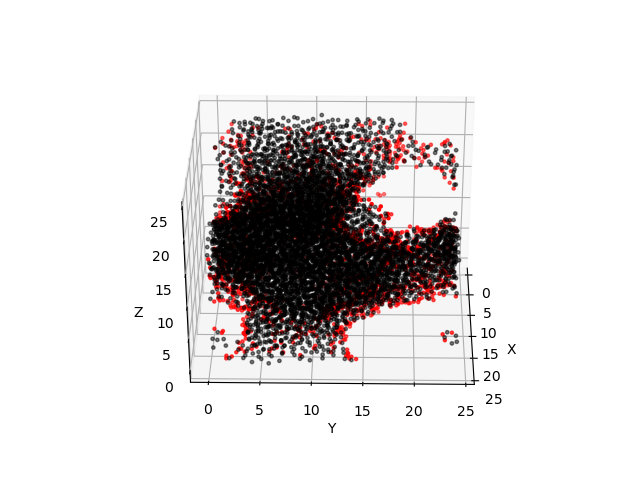}
\caption{N=15625}
\end{subfigure}
\begin{subfigure}[t]{0.3\textwidth}
\includegraphics[width=\textwidth]{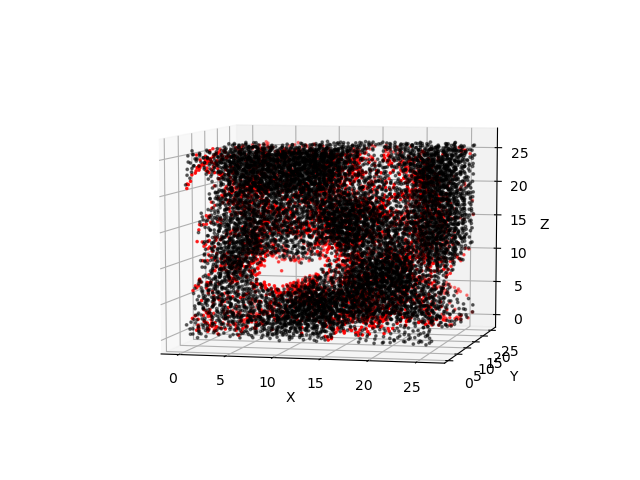}
\caption{N=19683}
\end{subfigure}
\bigskip
\begin{subfigure}[t]{0.3\textwidth}
\includegraphics[width=\textwidth]{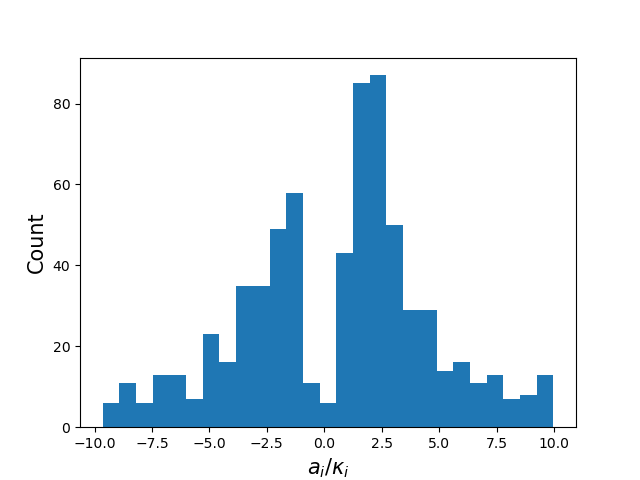}
\caption{N=8000}
\end{subfigure}
\begin{subfigure}[t]{0.3\textwidth}
\includegraphics[width=\textwidth]{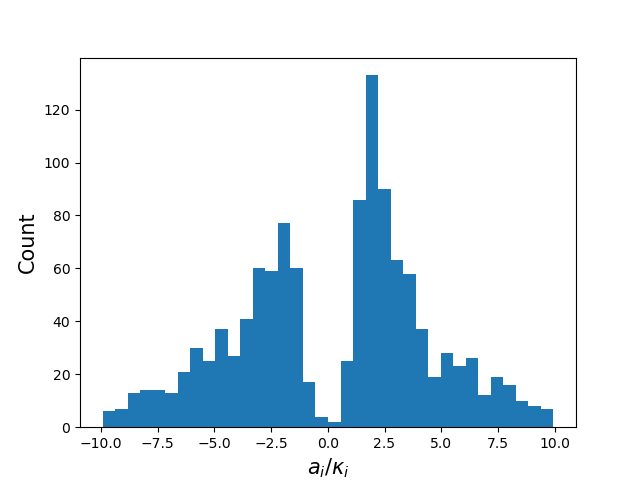}
\caption{N=15625}
\end{subfigure}
\begin{subfigure}[t]{0.3\textwidth}
\includegraphics[width=\textwidth]{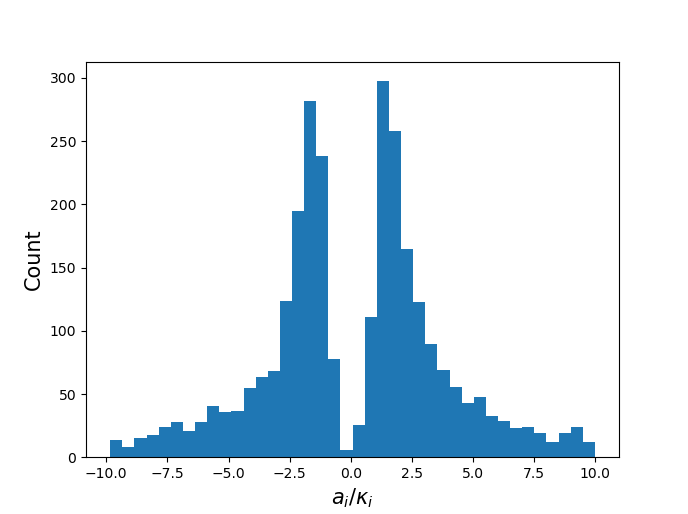}
\caption{N=19683}
\end{subfigure}
\caption{(a-c) Configuration space displaying liquid particles belonging to one state (black) against the interfacial particles generated to represent their surface (red) at $T^*=0.36$,$\rho^*=1.12$ for $N=8000,15625,19683$. Histograms (d-f) of $\frac{a_i}{\kappa_i}$ from Eq. \ref{eq:local} demonstrate that as system size increase, regions of positive and negative curvatures scaled by local surface area increase. This shift allows for consistent interfacial free energy values as $N$ increases. }
\label{fig:8}
\end{figure}

\begin{figure}[H]
\centering
\includegraphics[scale=0.4]{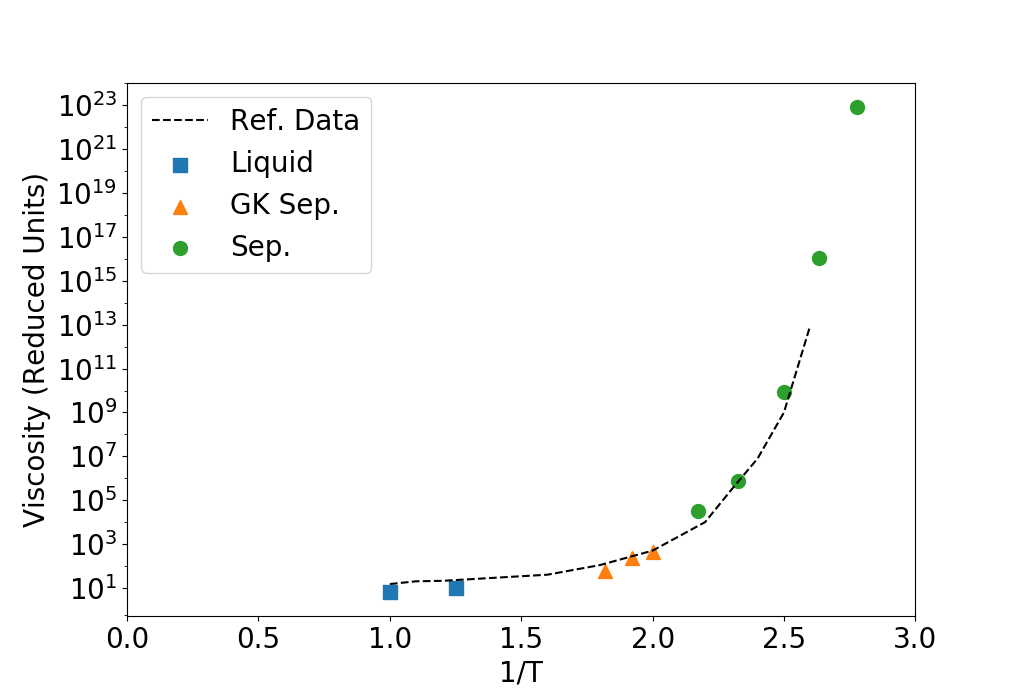}
\caption{Viscosity computed using the Markov Network Model(MNM) from Eq.(\ref{eq:two_state_2}). Viscosity values where Green-Kubo is still reliable within the liquid-liquid binodal are depicted by triangles. These systems are used to estimate the trial frequency $\nu_{0}$. Homogeneous liquid ones are denoted by squares, and those below the temperature limit of Green-Kubo where only MNM is reliable are shown by circles. The dashed line is from the data in Ref.\cite{Li}, where MNM is used with thousands of inherent structures as basins.}
\label{fig:9}
\end{figure}

Systems within the temperature range $T^{*}=[0.5,0.52,0.55]$ are still below the liquid-liquid transition line at $\rho^{*}=1.12$ but above the mode-coupling temperature $T^{*}=0.435$, hence their viscosities can be estimated using the Markov Network Model (Eq.\ref{eq:two_state_2}) and the conventional Green-Kubo method. The difference between these values was then adjusted by setting the trial frequency $\nu_0$, which was then used for all other temperatures. In conjunction with homogeneous liquid viscosities at $T^{*}=1.0,0.8$, the viscosities computed using our model are shown in~\ref{fig:9}. 

\begin{table}[H]
\centering
\begin{tabular}{|l|l|l|l|l|l|l|l|l|}
\hline
$T^*$ & 0.36 & 0.38 & 0.4 & 0.43 & 0.46 & 0.5 & 0.52 & 0.55 \\ \hline
$\Delta P^*$ & 0.14 & 0.12 & 0.11 & 0.091 & 0.07 & 0.062 & 0.06 & 0.05 \\ \hline
$\sum_i \overline{a}\frac{1}{\kappa_i}$ & 265 & 241 & 162 & 124 & 114 & 109 & 104 & 113 \\ \hline
\end{tabular}
\caption{Sum of curvature radius per-particle and pressure difference of liquid states for $\rho^*=1.12$ systems used to construct the reduced viscosity curve. The trial frequency $\nu_0^{-1}=10 s$ used at each temperature $T^*$ is the one used to match Green-Kubo data from $T^{*}=[0.5,0.52,0.55]$. Shear stress $\sigma_0$ for each system is of the order $10^{-3}$ in reduced pressure unit, and the dimensions of the simulation box is 24 reduced units. The average area $a$ occupied by each particle along the surface is $a=\pi r^2=\pi (0.98)^2=3.015$ estimated from the bulk density, which doesn't vary as a function of temperature. Using these data and Eq.\ref{eq:two_state_2} one can construct ~\ref{fig:9}. }
\label{table:4}
\end{table}

\section{Concluding remarks}

Using the WCN as an order parameter, we reproduced the gas-liquid binodal line and successfully extended the calculation to liquid-liquid binodal and its critical temperature for the 80:20 Kob-Andersen model. There is clear thermodynamic evidence for the two coexisting liquid states based on their pressure and density profiles. The coexistence densities of liquid states quenched at different bulk densities obey the lever rule at a particular temperature, which is a strong indication of the local thermodynamic equilibrium. 

As the temperature decreases to the glass transition region, the Markov Network Model is adopted for the computation of temperature dependent viscosity using the liquid-liquid phase separation picture. The viscosity increases by almost 16 orders of magnitude over the temperature range of 0.55 to 0.36, clearly indicating super-Arrhenius behavior. The physical mechanism for such a drastic dynamical slowing down is the coarsening kinetics of the liquid-liquid phase separation as indicated by the Markov Network Model calculations. It is demonstrated that the coarsening dynamics of the liuqid-liquid phase separation may serve as a viable picture for the understanding of glass transitions for the studied potential.

Application to the ST2 water model~\cite{Stillinger} also indicates liquid-liquid phase separation, as indirectly demonstrated by other methods ~\cite{sciortino2011study,yagasaki,Liu}, and the temperature-dependent viscosity calculations are under way. The liquid-liquid separation appears in many different water models~\cite{doi:10.1073/pnas.2526573123, 10.1063/5.0196964, 10.1063/5.0099520}, for which we plan to classify and analyze these systems in a similar fashion as detailed in this paper, and will be presented in a future paper. 
If  super-Arrhenius behavior is also observed, then the liquid-liquid phase separation mechanism may provide a simple picture for understanding glass transitions in general. Naturally, it will be interesting to explore other salient features of glass transition,  as shown in other glass transition studies~\cite{wolynes_book} within the current liquid-liquid phase separation picture.

 \section{Acknowledgement}

This work was supported by the Division of Chemical and Biological Sciences, Office of Basic Energy Sciences, U.S. Department of Energy, under Contact No. DE-AC02-07CH11358 at Iowa State University. 

\appendix
\section{Local Curvature Calculations}
\label{sec:local_curvature}

Two approaches are utilized to estimate the local surface of the interfacial particles for local curvature estimation to ensure that reliable results are obtained consistently.

The first approach utilizes local neighbor lists for interfacial particles, gathered under the criterion that $70\%$ of a particle's neighbors belong to another state. Midpoints between each of these tagged particles are made with their neighbors belonging to the other state, generating surfaces with many more points to sample than just the particles along the surface, which is limited by the density. This approach creates a surface with many more particles to sample along the liquid-liquid interface, allowing for a more precise fitting of local surfaces. 

The second approach uses log-likelihood information from the PC space to determine which particles have the highest probability of belonging to either state. The issue with this approach is that K-means clustering in the PC space is non-deterministic; therefore,  the particles assigned to the surface may vary, and the number of particles collected using this approach is far fewer than that of the prior method. Additionally, the log-likelihood gap between states decreases as the liquid states become more thermodynamically similar close to $T^*_c$, making the first approach more reliable and the method used in all analysis. 

\begin{figure}
\includegraphics[scale=0.8]{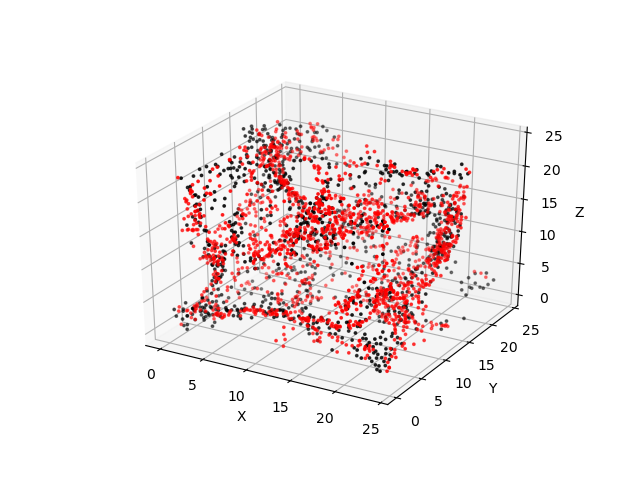}
\parbox{6in}
\centering
{
\caption{Comparison of methods for construction of local surface that embeds an interfacial particle, using neighbor lists (red) and using PC log-likelihood (black). Log-likelihood gets more inconsistent closer to the liquid-liquid binodal line, and so neighbor lists are more reliable for local surface construction.  }
}
\label{fig:10}
\end{figure}

Once a local surface containing an interfacial particle is constructed, the following procedure leads to local curvature:
\begin{enumerate}
	\item Generate surface point neighbor lists (from PC's or neighbor information).
	\item Points and their neighbors were fitted to the 2D quadric equation $F(x,y,z)=ax^2 + by^2 + cz^2 + 2exy + 2fyz + 2gxz + 2lx + 2my + 2nz + d =0$ using the least squares method. We did not wish to calculate the nearest point on the plane fitted to the surface point of interest. Instead, we assume $F(P)=0$, where $P$ is the surface point of interest. 
	\item Using coefficients from fitting, compute first fundamentals of $F(x,y,z)$,
		
		\begin{equation}
		E = 1 + \frac{F_x^2}{F_z^2} , \quad F = \frac{F_xF_y}{F_z^2} , \quad G = 1 + \frac{F_y^2}{F_z^2}
		\end{equation}			
		
		and the second fundamentals 
		
		\small
		\begin{equation}
		L = \frac{1}{F_z^2 |\nabla F|} \begin{vmatrix} F_xx & F_xz & F_x \\ F_zx & F_zz & F_z \\ F_x & F_z & 0 \end{vmatrix}, \\
		M= \frac{1}{F_z^2 |\nabla F|} \begin{vmatrix} F_xy & F_yz & F_y \\ F_zx & F_zz & F_z \\ F_x & F_z & 0 \end{vmatrix}, N = \frac{1}{F_z^2 |\nabla F|} \begin{vmatrix} F_yy & F_yz & F_y \\ F_zy & F_zz & F_z \\ F_y & F_z & 0 \end{vmatrix}.
		\end{equation}
		\normalsize

\end{enumerate}

Once the fundamentals have been computed, the principal curvatures can be determined by constructing matrices 

\begin{equation}
A = \begin{bmatrix} L & M \\ M & N \end{bmatrix} , \quad B = \begin{bmatrix} E & F \\ F & G \end{bmatrix}.
\end{equation}
The eigenvalues of $B^{-1}A$ correspond to $k_1, k_2$, with mean curvature $H=(k_1 + k_2)/2$. 

\begin{figure}[H]
\centering
\resizebox{\columnwidth}{!}
{
\begin{subfigure}{0.45\textwidth}
\includegraphics[width=2.3in]{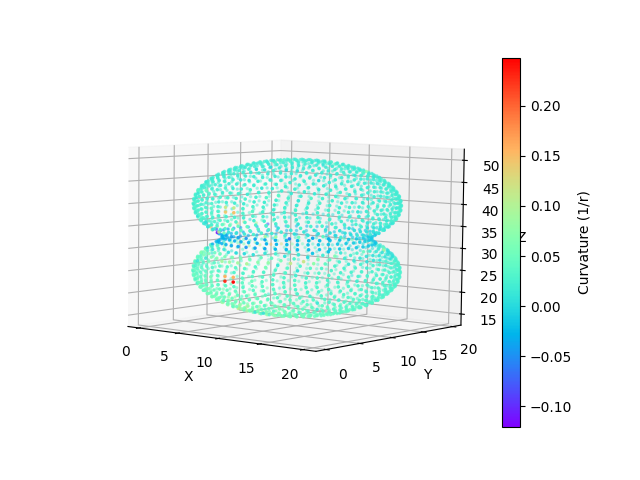}
\caption{Diatomic}
\end{subfigure}
\hfill
\begin{subfigure}{0.45\textwidth}
\includegraphics[width=2.3in]{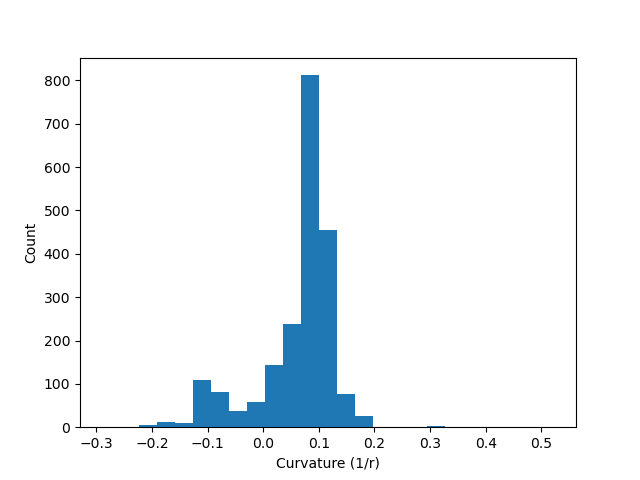}
\caption{Histogram}
\end{subfigure}
}
\caption{ (a) Curvature estimation of a diatomic molecule using 2D quadric plane fit and (b) histogram of curvature values. The hemispheres of the diatomic have radius $r=10$, resulting in curvature values consistent with $1/r$. This verification of accuracy of a surface with both convex and concave curvature allows us to extend it to the Kob-Andersen case.}
\label{fig:11}
\end{figure}

\begin{figure}[H]
\centering
\resizebox{\columnwidth}{!}
{
\begin{subfigure}{0.45\textwidth}
\includegraphics[width=2.3in]{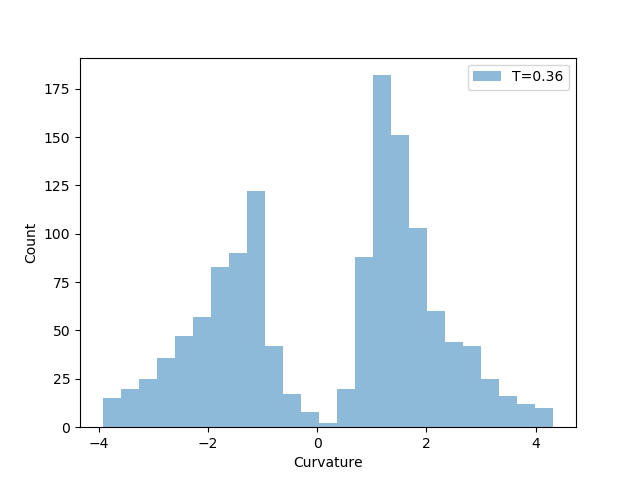}
\caption{Histogram data}
\end{subfigure}
\hfill
\begin{subfigure}{0.45\textwidth}
\includegraphics[width=2.3in]{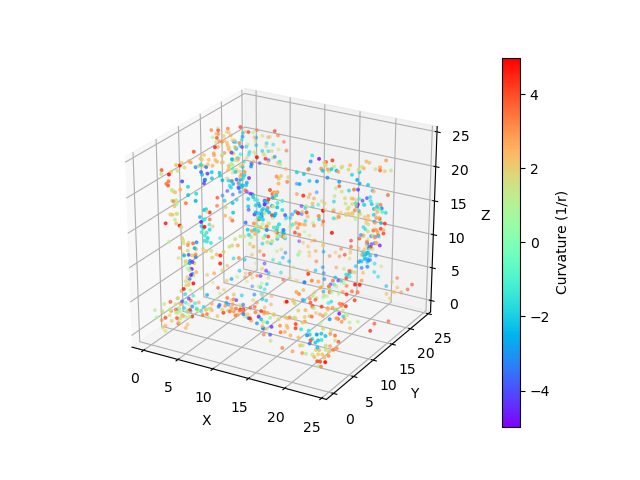}
\caption{Color mapping of surface particles}
\end{subfigure}
}
\caption{ Curvature data collected using PC space log-likelihood to detect interfacial particles of $T^{*}=0.36$, $\rho^{*}=1.12$ Kob-Andersen system. 2D quadric plane fit was used to calculate local curvature for each interfacial particle. }
\label{fig:12}
\end{figure}

To test our algorithm,  the curvatures of a perfect sphere with $r=1$ and a diatomic molecule with atomic radius $r=10$ and bond length=2. The expected curvature values for the sphere and hemispheres of the diatoms are given by the relation $\kappa = \frac{1}{r}$.  Figure \ref{fig:11} shows the diatomic curvature estimations. \ref{fig:12} shows the curvature distribution of the interfacial particles under a particular liquid-liquid phase separation condition.

\section{State dynamics}
\label{sec:wcn-dynamics}

\begin{figure}[H]
\centering
\includegraphics[scale=0.55]{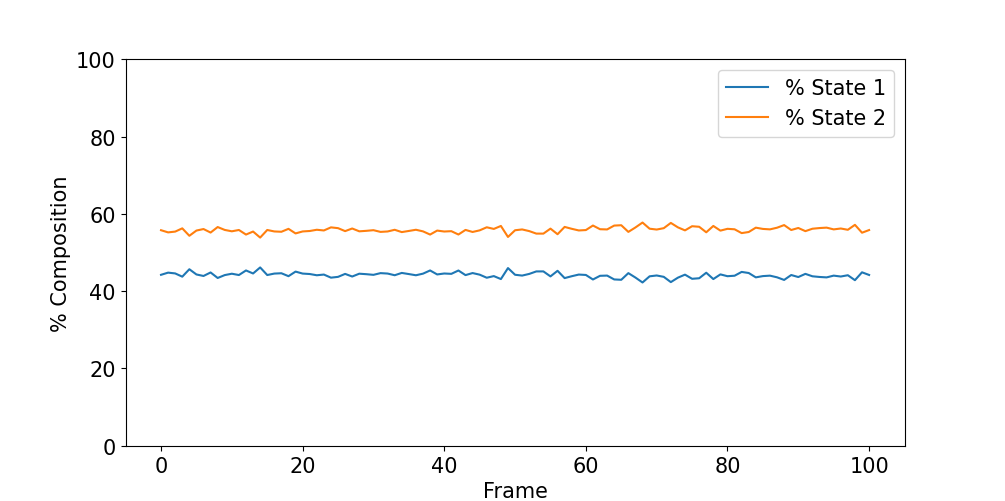}
\parbox{6in}
{
\caption{Time evolution of WCN classification for $T^*=0.36$, $\rho^*=1.12$ system over the course of 5ns after equilibration for 2.4$\mu$s, which is below the standard equilibration time of 3.6$\mu$s for which all relevant data is collected. Perfcent state composition of the box is demonstrated to be stable within the given data collection window and matches composition values after 3.6$\mu$s.  }
}
\label{fig:13}
\end{figure}

\begin{figure}[H]
\centering
\begin{subfigure}[t]{0.3\textwidth}
\includegraphics[width=\textwidth]{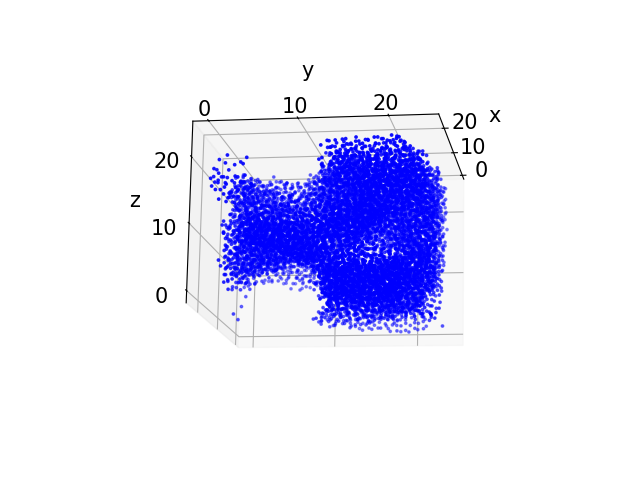}
\caption{Quench, t=600ns, 67\%}
\end{subfigure}
\begin{subfigure}[t]{0.3\textwidth}
\includegraphics[width=\textwidth]{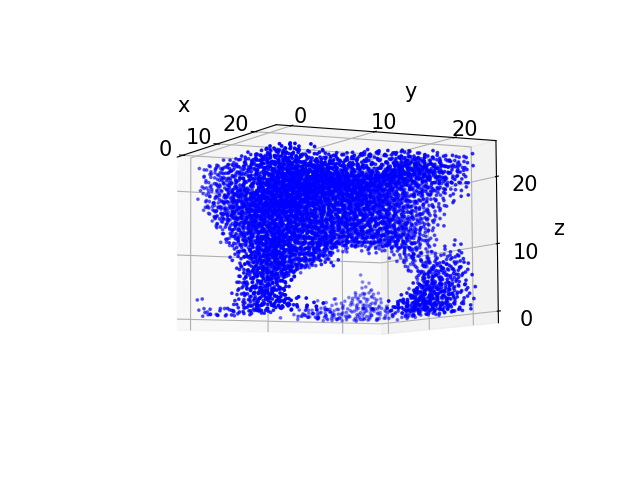}
\caption{Quench, t=2.4$\mu$s, 59\%}
\end{subfigure}
\begin{subfigure}[t]{0.3\textwidth}
\includegraphics[width=\textwidth]{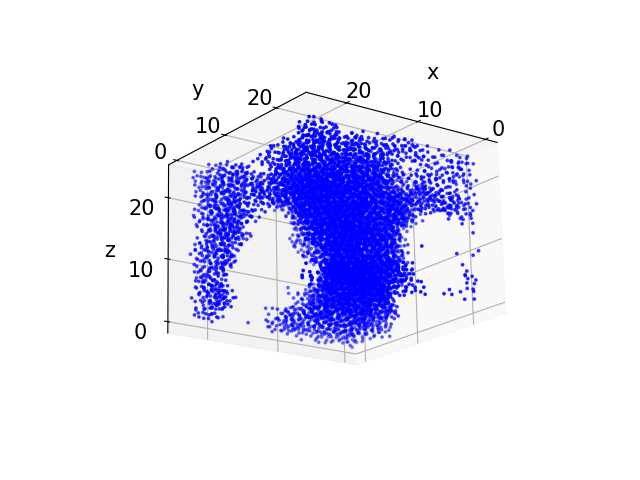}
\caption{Cool, t=360ns, 60\%}
\end{subfigure}
\caption{Configuration space displaying liquid particles belonging to one state (blue) for an NPT system at N=15625, P=0, T*=0.3 with initial temperature T*=1.2. Pathway dependence of configurational morphology for a quenched system (a) equilibrated for 600ns and (b) for 2.4$\mu$s. (c) The same T*=1.2 system was cooled to T*=0.3 in 2.37x10$^{10}$K/s and equilibrated for 360ns. Both the cooling rate system and quenched system display similar morphologies and percent compositions.}
\label{fig:8}
\end{figure}

\section{Block Average Analysis~\cite{Frenkel-Smit}}
\label{sec:block}

\begin{figure}[H]
\centering
\includegraphics[scale=0.6]{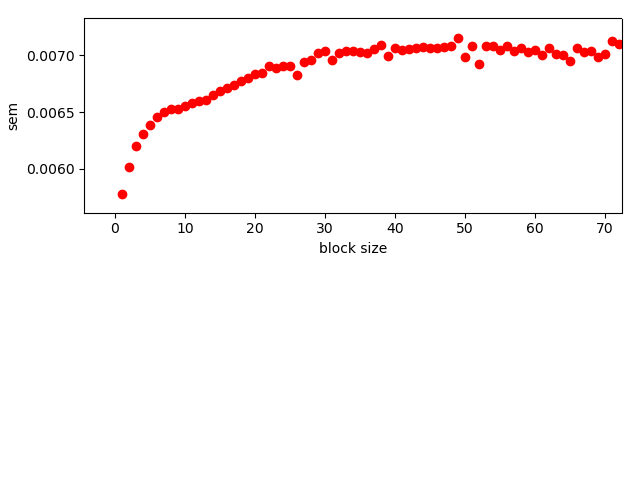}
\parbox{6in}
{
\caption{Block average of diagonal pressure tensor components for 200 particles at $T^*=1.0$, $\rho^* = 1.12$. Block size is modulated by number of particles sampled. High thermal fluctuations allow very good sampling of thermodynamic data. This serves as the ideal case for sampling small number of particles within a given region of the simulation box.  }
}
\label{fig:15}
\end{figure}

\begin{figure}[H]
\centering
\includegraphics[scale=0.6]{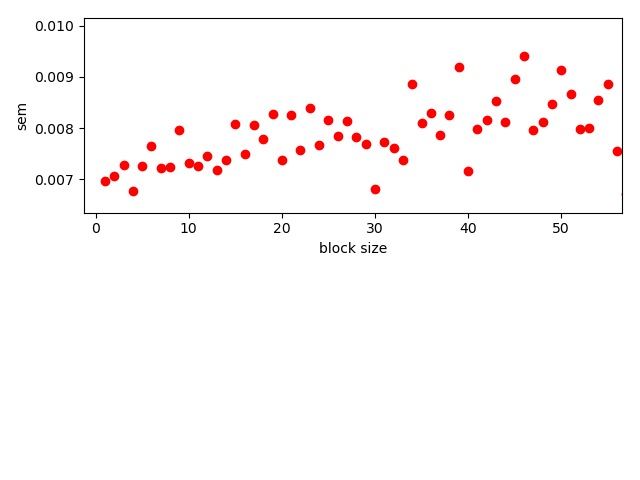}
\parbox{6in}
{
\caption{Block average of normal pressure tensor $P_N(r)$ for particles within a thin radial shell along {\it convex} liquid-liquid interface at $T^*=0.3$, $\rho^* = 1.12$. Block size is modulated by number of particles sampled. For low termperature systems, dynamic constraints near the glass transition prevent well-behaved pressure fluctuations within a small region of the simulation box. While fluctuations are high as block size increases, long averages nontheless plateau and yield statistically uncorrelated pressure values with well-behaved means and standard deviations. }
}
\label{fig:16}
\end{figure}

\begin{figure}[H]
\centering
\includegraphics[scale=0.6]{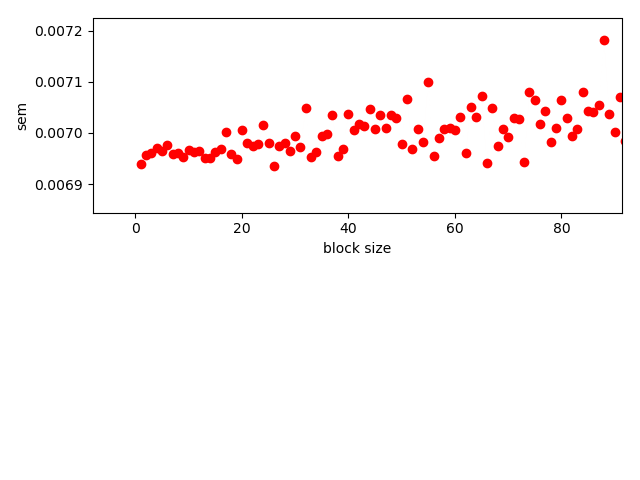}
\parbox{6in}
{
\caption{Block average of normal pressure tensor $P_N(r)$ for particles at within a thin radial shell along {\it concave} liquid-liquid interface $T^*=0.3$, $\rho^* = 1.12$. Block size is modulated by number of particles sampled. For low termperature systems, dynamic constraints near the glass transition prevent well-behaved pressure fluctuations within a small region of the simulation box. While fluctuations are high as block size increases, long averages nontheless plateau and yield statistically uncorrelated pressure values with well-behaved means and standard deviations. }
}
\label{fig:17}
\end{figure}

\bibliographystyle{unsrt}
\bibliography{references}

\end{document}